# Low-Energy Electron Microscopy Studies of
# Interlayer Mass Transport Kinetics on TiN(111)

S. Kodambaka,*,a Navot Israeli,[1,b] J. Bareño,a W. Święch,a
Kenji Ohmori,a I. Petrov,a and J.E. Greenea,b

aDepartment of Materials Science and the Frederick Seitz Materials Research Laboratory
University of Illinois, 104 South Goodwin Avenue, Urbana, IL 61801 USA
bDepartment of Physics, University of Illinois, 1110 West Green St., Urbana, IL 61801 USA

## Abstract

In situ low-energy electron microscopy was used to study interlayer mass transport kinetics during annealing of three-dimensional (3D) TiN(111) mounds, consisting of stacked 2D islands, at temperatures T between 1550 and 1700 K. At each T, the islands decay at a constant rate, irrespective of their initial position in the mounds, indicating that mass is not conserved locally. From temperature-dependent island decay rates, we obtain an activation energy of 2.8±0.3 eV. This is consistent with the detachment-limited decay of 2D TiN islands on atomically-flat TiN(111) terraces [Phys. Rev. Lett. 89 (2002) 176102], but significantly smaller than the value, 4.5±0.2 eV, obtained for bulk-diffusion-limited spiral step growth [Nature 429, 49 (2004)]. We model the process based upon step flow, while accounting for step-step interactions, step permeability, and bulk mass transport. The results show that TiN(111) steps are highly permeable and exhibit strong repulsive temperature-dependent step-step interactions that vary between 0.03 and 0.76 eV-Å. The rate-limiting process controlling TiN(111) mound decay is surface, rather than bulk, diffusion in the detachment-limited regime.

---

*Corresponding author. Address: ESB 1-135, 1101 W. Springfield Avenue, Urbana, IL 61801 USA.
Tel.: +1-217-3333481; fax: +1-217-2441638. E-mail address: kodambaka@mrl.uiuc.edu
[1] Present address: Department of Physics of Complex Systems, Weizmann Institute of Science, Rehovot, 76100, Israel.

## 1. Introduction

NaCl-structure TiN and related transition-metal (TM) nitrides are widely used as hard wear-resistant coatings on cutting tools, diffusion-barriers in microelectronic devices, corrosion-resistant layers on mechanical components, and abrasion-resistant thin films on optics and architectural glass. Controlling the microstructural and surface morphological evolution of polycrystalline TM nitride films is important in all of these applications. This fact has spurred interest in modeling polycrystalline TM nitride thin film growth [1], a complex phenomenon controlled by the interplay of thermodynamic driving forces and kinetic limitations, as a function of deposition conditions. Developing a quantitative model, however, requires knowledge of atomic-level processes, site-specific surface energetics, and rate-limiting mechanisms. Recently, considerable progress has been made toward the determination of absolute orientation-dependent step energies, step stiffnesses, and kink energies [2,3], as well as identifying the mechanisms controlling the kinetics of two-dimensional (2D) TiN island coarsening/decay (Ostwald ripening) [4,5] and island shape equilibration on (001) and (111) TiN terraces [6-8]. Here, we focus on the decay kinetics of 3D TiN(111) mounds, consisting of stacked 2D islands, in order to probe interlayer mass transport mechanisms, step-step interactions, and step permeability on TiN(111) at elevated temperatures.

Experimental studies of the coarsening/decay kinetics of 2D homoepitaxial semiconductor and fcc metal islands stacked in 3D mound geometries indicate that the rate-limiting processes controlling island decay behavior are both qualitatively and quantitatively different from those of isolated islands on terraces [9-13]. Novel interlayer mass transport mechanisms, including the concerted downward motion of adatoms and site-selective adatom descent have been proposed to explain the observed behavior [14,15]. Bartelt and co-workers [16,17] and, more recently,

Ondrejcek *et al*. [18] have found that bulk, rather than surface, mass transport controls surface dynamics at elevated temperatures. The coarsening/decay behavior of 3D mounds, consisting of layer structures in a "wedding cake" configuration, has been studied extensively, both experimentally [9,12,16] and theoretically [19-21].

In this paper, we present the results of in situ high-temperature (T = 1550-1700 K) low-energy electron microscopy (LEEM) [22] measurements of the coarsening/decay kinetics of concentrically-stacked 2D TiN adatom islands on TiN(111) terraces. Island areas decrease with annealing time at a constant rate, irrespective of their initial position, indicating that mass is not conserved locally. From temperature-dependent island decay rates, we obtain an activation energy $E_d$ of 2.8±0.3 eV. This is significantly lower than the value, 4.5±0.2 eV, for bulk-diffusion-limited growth of TiN(111) spiral steps [23]. Based upon these results in combination with a step-flow model [20,21], while accounting for step-step interactions, step permeability (the ability for adatoms to migrate across steps without attaching to them [24]), and bulk mass transport, we determine that TiN(111) steps exhibit repulsive step-step interactions and that the TiN(111) mounds decay via surface mass transport in the detachment-limited regime combined with high step permeability rates.

## 2. Experimental procedure

Epitaxial TiN(111) layers, 2000-Å-thick, were grown on polished $Al_2O_3(0001)$ substrates (0.5-mm-thick x 9 mm diameter) at $T_s$ = 1050 K in a load-locked multi-chamber ultra-high vacuum (UHV) system using magnetically-unbalanced dc magnetron sputter deposition [25] following the procedure described in Ref. 3. The back sides of the substrates had previously been coated with 1-μm-thick TiN layers to allow electron-beam heating without substrate decomposition. The TiN(111) samples were transferred to a UHV multichamber LEEM system [26], with a base pressure of $2 \times 10^{-10}$ Torr, which is equipped with facilities for residual gas analysis (RGA), electron-beam evaporation, ion sputtering, Auger electron spectroscopy (AES), and low-energy electron diffraction (LEED). Sample temperatures were measured by optical pyrometry and calibrated using temperature-dependent TiN emissivity data obtained by spectroscopic ellipsometry. The TiN layers were degassed in the LEEM sample preparation chamber at 1073 K for approximately 2 h. This procedure results in sharp 1x1 LEED patterns with a 3-fold symmetry as expected for bulk-terminated TiN(111) [3]. In-situ AES analyses indicate that the samples contain $\simeq$ 3 mole% oxygen, most likely in the form of TiO which is isostructural [27] and mutually soluble with TiN.

Homoepitaxial TiN(111) overlayers, 50-200 Å thick, were deposited at 1023 K by reactive evaporation from Ti rods (99.999% purity) at a rate of $\simeq$ 0.02 ML/s and annealed for 2-3 days in $5 \times 10^{-8}$ Torr $N_2$ (99.999%) at temperatures T > 1200 K. The deposition/annealing cycles were repeated until large (> 1000 Å) atomically-smooth TiN(111) terraces and 3D mounds, consisting of concentrically-stacked 2D TiN islands, are obtained.

Bright-field (BF) LEEM images of the coarsening/decay of these islands, whose average areas range from $2 \times 10^{-2}$ to $13 \times 10^{-2}$ $\mu m^2$, were acquired at a video rate of 30 frames/s as a function of annealing time and temperature. Pixel resolutions varied between $\simeq$ 55 and 85 Å for fields of view ranging from 2.6 to 4 $\mu m$. Typical electron probe beam energies were 5 to 25 eV. The samples were allowed to thermally stabilize at each temperature for 10 to 15 s prior to acquiring LEEM videos. From each measurement sequence, time-dependent island boundaries and the total areas A of each layer in the mounds were determined using Image SXM, an image

processing software.[2]

## 3. Results and discussion

Fig. 1 shows BF-LEEM images (fields of view = 8 μm) of the surface morphologies of: (a) a sputter-deposited heteroepitaxial TiN(111) layer, and (b) and (c) different regions of the same sample following two deposition/annealing cycles. The thick dark lines visible in Fig. 1a are $\simeq$ 300-Å-deep grooves bounding single-crystalline TiN(111) domains oriented $180^o$ with respect to each other due to the three-fold symmetry of the TiN(111) surface. Bilayer-height TiN(111) islands[3] with truncated-hexagonal shapes bounded by alternating long and short $\langle 110 \rangle$ steps [3] are shown in Fig. 1b, while Fig. 1c is an image of a single domain with $\simeq$ 1500-Å-wide atomically-smooth terraces separated by bilayer-height steps. Sample surfaces corresponding to Figs. 1b and 1c are typical of those used in subsequent annealing experiments in which we follow coarsening/decay of 3D mounds.

Figs. 2a-c are representative BF-LEEM images (4 μm fields of view) obtained from a TiN(111) sample during annealing at T = 1552 K. The images, part of a 30 frame/s video file, were acquired at times t = 0, 95, and 230 s. (We define t = 0 as the time at which the first image is acquired.) During annealing, we observe dissolution of the TiN(111) islands which retain their truncated-hexagonal shape during the process. Fig. 2d is a plot of area A vs. annealing time t for the islands labeled 1, 2, 3, and 4 in Figs. 2a-2c. Islands 1 and 3 are situated on top of islands 2 and 4, respectively. We find that the island areas decrease with time at a rate dA/dt, which depends both on the island geometry and the local environment. For example, islands 1 and 2,

---

[2] Image SXM, developed by Prof. Steve Barrett, Surface Science Research Centre, Liverpool, England, 2002 (http://reg.ssci.liv.ac.uk).
[3] The [111] direction in NaCl-structure TiN is polar, consisting of alternating layers of Ti and N atoms.

which have similar geometry, decay at 6.68±0.07x10$^4$ and 6.25±0.08x10$^4$ Å$^2$/s, respectively, while islands 3 and 4, located in a different mound, decay at 10.18±0.06x10$^4$ and 9.66±0.08x10$^4$ Å$^2$/s. These results are typical of LEEM data obtained for over 50 islands at all temperatures in the range 1550-1700 K.

In order to isolate the rate-limiting mechanisms controlling TiN(111) island decay kinetics at elevated temperatures, we analyzed the decay behavior of 3D mounds consisting of approximately concentrically-stacked island structures. Fig. 3a is a typical BF-LEEM image acquired during annealing a TiN(111) sample at T = 1559 K. We follow the time- and temperature-dependent decay kinetics of several successive layers in the circled region shown at higher magnification in Fig. 3b. This simple configuration allows us to apply an available step flow model [21], developed following the general approach of Burton-Cabrera-Frank (BCF) [28], which incorporates step-step interactions, step permeability, and both surface and bulk mass transport.

Figs. 3c-3f are plots of island area A vs. annealing time t for four successive layers at: (c) T = 1559, (d) 1590, (e) 1622, and (f) 1651 K. Note the absence of any significant increase in the areas of lower layer islands during the decay of upper layers. This is the signature of (1) strong step-step repulsion leading to a very stiff step system and/or (2) the absence of local mass conservation. Of the three possible mass transport mechanisms -- adatom terrace diffusion across highly permeable steps, exchange with the bulk, and evaporation -- that can lead to non-conservation of mass during annealing, we rule out evaporation since TiN is a highly refractory compound (melting point $T_m$ = 3200 K and cohesive energy ≃ 14 eV) with desorption energies of 8.8 and 10 eV for TiN and Ti adspecies, respectively, from N-terminated TiN(111) [29].

In the following paragraphs, we describe a step-flow model that accounts for step-step interactions, step permeability, and bulk-mass transport processes. We calculate time-dependent changes in island radii for each of three limiting cases: (1) island decay via surface mass transport across impermeable steps in the absence of bulk mass transport, (2) surface mass transport across permeable steps, and (2) both bulk and surface mass transport across impermeable steps. We then compare the calculated results with experimental data in order to determine the dominant mechanism.

In our model, the $i^{th}$ island of area A$_i$ in a given 3D mound is characterized by its average radius $r_i$, defined as $r_i = \sqrt{A_i/\pi}$ . ($i$ is a running index which increases from the top of the mound to the bottom.) In the absence of net mass change due to deposition, evaporation, and bulk diffusion, the adatom concentration fields between the islands are described by the 2D steady-state diffusion equation

$$\nabla^2 \rho_i(r) = 0, \text{ with } r_{i-1} \le r \le r_i, \tag{1}$$

where $\rho_i(\text{r})$ is the adatom concentration on the $i^{th}$ terrace. A steady-state solution is justified since the time scales associated with equilibration of adatom concentration fluctuations on the terraces are much shorter than those associated with island step motion. We solve Eq. (1) using boundary conditions (given below) which specify adatom fluxes into (or out of) the islands and, hence, determine the rate of change of island radii. Assuming first-order kinetics, the flux boundary conditions at the $i^{th}$ island are

$$
\begin{aligned}
D_s \left. \frac{\partial \rho_i}{\partial r} \right|_{r_i} &= K_d \left( \rho_i|_{r_i} - \rho_i^{eq} \right) + p \left( \rho_i|_{r_i} - \rho_{i-1}|_{r_i} \right), \\
D_s \left. \frac{\partial \rho_{i-1}}{\partial r} \right|_{r_i} &= -K_d \left( \rho_{i-1}|_{r_i} - \rho_i^{eq} \right) + p \left( \rho_i|_{r_i} - \rho_{i-1}|_{r_i} \right).
\end{aligned}
\tag{2}
$$

In Eq. (2), $D_s$ is the surface diffusivity, $K_d$ is the attachment/detachment rate, and $p$ is the step permeability. $\rho_i^{eq}$ is the equilibrium adatom concentration in the vicinity of the $i^{th}$ step, which is related to the step chemical potential $\mu_i$ and the equilibrium adatom concentration $\rho_\infty^{eq}$ at a straight isolated step through the Gibbs-Thomson relation [4],

$$\rho_i^{eq} = \rho_\infty^{eq} \cdot \exp\left(\frac{\mu_i}{kT}\right) \approx \rho_\infty^{eq} \cdot \left(1 + \frac{\mu_i}{kT}\right). \tag{3}$$

In deriving Eq. (3), we have assumed (and provide justification below) that $\left(\dfrac{\mu_i}{kT}\right) \ll 1$.

For an island in a stack, $\mu_i$ depends on the island curvature and on elastic and entropic repulsive interactions between nearest-neighbor steps, accounting for which yields the relationship[4] [9,21,30]

$$\mu_i = \frac{\Omega B}{r_i} + \Omega G\left(\frac{2r_{i+1}}{(r_{i+1}+r_i)(r_{i+1}-r_i)^3} - \frac{2r_{i-1}}{(r_i+r_{i-1})(r_i-r_{i-1})^3}\right). \tag{4}$$

In Eq. (4), the parameter B defines the energy scale and is a measure of the surface equilibrium chemical potential [2], $G$ is the step-step repulsive interaction strength, and $\Omega$ is the unit TiN molecular area. In case of an isolated island on a terrace, Eq. (4) reduces to $\mu = B\Omega/r$, which is valid for anisotropic islands with *any* arbitrary equilibrium shape and can be used to model island decay kinetics using circular islands having the same area [2]. Since $B = 0.23$ eV/Å [3], $\Omega$ = 7.2 Å$^2$ for TiN(111), and the average areas of islands analyzed here are $> 2 \times 10^{-2}$ μm$^2$, we find that $(\mu_i/kT) \leq 0.01$ at T $\geq$ 1550 K. This validates the condition, $(\mu_i/kT) \ll 1$, required for Eq. (3).

---

[4] In deriving Eq. (4), we assume circular-shaped islands that are concentrically stacked and separated by a distance that is smaller than the island radii.

Solving Eq. (1) for $\rho_i$, with the boundary conditions specified by Eqs. (2)-(4), yields the rate of change $dr_i/dt$ of the island radius. For the limiting case in which the steps are impermeable, i.e. $p = 0$, and there is no bulk mass transport, we obtain

$$\frac{dr_i}{dt} = \frac{\Omega K_d \rho_\infty^{eq}}{r_i\, kT}\left[\frac{\mu_i - \mu_{i+1}}{\dfrac{K_d}{D_s}\ln\left(\dfrac{r_i}{r_{i+1}}\right) - \left(\dfrac{1}{r_i} + \dfrac{1}{r_{i+1}}\right)} - \frac{\mu_{i-1} - \mu_i}{\dfrac{K_d}{D_s}\ln\left(\dfrac{r_{i-1}}{r_i}\right) - \left(\dfrac{1}{r_{i-1}} + \dfrac{1}{r_i}\right)}\right]. \tag{5}$$

If the steps are permeable ($p > 0$), however, adatoms can hop across without becoming incorporated at the step edges and mass is not conserved locally. This results in coupling of the adatom diffusion fields on all terraces. Consequently, the equations describing the areal rate of change for the $i^{th}$ island are linked with those describing each of the other islands in the stack and we solve the full equation set.

Adatom transport between the bulk and the surface also leads to local non-conservation of mass. In order to account for this possibility, we follow Ref. [16] and assume that mass exchange with the bulk occurs only near island edges at a rate $K_{bulk}$. The adatom flux $J_i$ from the $i^{th}$ step to the bulk is then given by

$$J_i = K_{bulk}\left(\frac{\rho_\infty^{eq}}{kT}\right)\mu_i, \tag{6}$$

where we make the assumption that the bulk chemical potential is at equilibrium. Combining Eqs. (5) and (6), we obtain an expression for the velocity $dr_i/dt$ of impermeable steps ($p = 0$) in the presence of bulk diffusion,

$$\frac{dr_i}{dt} = \frac{\Omega K_d \rho_\infty^{eq}}{kT \cdot r_i}\left[\frac{\mu_i - \mu_{i+1}}{\dfrac{K_d}{D_s}\ln\left(\dfrac{r_i}{r_{i+1}}\right) - \left(\dfrac{1}{r_i} + \dfrac{1}{r_{i+1}}\right)} - \frac{\mu_{i-1} - \mu_i}{\dfrac{K_d}{D_s}\ln\left(\dfrac{r_{i-1}}{r_i}\right) - \left(\dfrac{1}{r_{i-1}} + \dfrac{1}{r_i}\right)} - K_{bulk} \cdot r_i \cdot \mu_i\right]. \tag{7}$$

The step-flow model outlined above contains several material parameters: $\Omega$, $D_s$, $K_d$, $K_{bulk}$, $p$, $\rho_\infty^{eq}$, $B$, and $G$, of which only $\Omega$ and $B$ are known for TiN(111). There are four independent variables which control island decay kinetics. The length scale $l = \dfrac{D_s}{K_d}$ defines the limiting surface mass transport mechanism. In the diffusion-limited regime, $l << \Delta x$, where $\Delta x$ is the average terrace width, while $l >> \Delta x$ in the detachment-limited regime. The ratios $\dfrac{p}{K_d}$ and $\dfrac{K_{bulk}}{K_d}$ describe the relative importance of step permeability and surface mass exchange with the bulk, respectively. Finally, the dimensionless quantity $g = \left(\dfrac{kT}{\Omega}\right)^2 \dfrac{G}{B^3}$ is a measure of the step-step interaction strength $G$. $\rho_\infty^{eq}$, the only term not included in these four parameters, only affects the time scale of step motion and can easily be accounted for by rescaling the time unit.

In defining the initial conditions, we set the total number of islands in the stack to be 20. Experimentally-measured initial radii $r_i(t = 0)$ of the top 4-7 islands (see, for example Fig. 4), together with the radii of the remaining islands, chosen such that they are equidistant from each other, are used as input to the model. (This choice of the total number of islands and the inter-island distances, while arbitrary, has no significant effect on calculated results.) Beginning with an initial set of $D_s/K_d$, $p/K_d$, and $K_{bulk}/K_d$ values, we first calculate time-dependent radii $r_i(t)$ for all the islands in the stack over time periods corresponding to the decay of five or more islands. Simulation times are then rescaled such that the calculated decay time for the top island matches that of the experiment. Finally, we compare the resulting island area trajectories with those of the LEEM data. This same self-consistent procedure is used for exploring the agreement at various points in parameter space by choosing different values of $D_s/K_d$, $p/K_d$, and $K_{bulk}/K_d$. We consider the three limiting cases, in which step permeability and bulk mass transport are absent, i.e. $p/K_d = K_{bulk}/K_d = 0$ and either step permeability or bulk mass transport is separately operative, i.e. $p/K_d > 0$ with $K_{bulk}/K_d = 0$ and $p/K_d = 0$ with $K_{bulk}/K_d > 0$. For each such set of results defined by ($D_s/K_d$, $p/K_d$, $K_{bulk}/K_d$), a minimization algorithm is used to find the value of g that minimizes any discrepancy between calculated and experimental data. Overall best fit solutions are then determined.

Fig. 4 shows a typical example of the agreement between experimental results at T = 1622 K and model predictions. The open circles in Figs. 4a and 4b are measured TiN(111) island radii as a function of time. The dashed lines in Fig. 4a are best fit calculated curves, obtained under the constraint of local mass conservation, i.e. $p = 0$ and $K_{bulk} = 0$, with $D_s/K_d = 0.1$ μm and $g = 0.0631$, for each island. While this relatively strong step-step interaction tends to minimize the recoil (the spike in lower island radii observed at times corresponding to the complete disappearance of the top island), the quality of the fits are far from satisfactory. As expected, imposing complete local mass conservation cannot explain the observed results, which require the presence of highly permeable steps and/or bulk mass transport.

The solid lines in Figs. 4a and 4b are calculated best fit solutions obtained by relaxing the local mass conservation constraint. We allow step permeability ($p > 0$) in the absence of bulk mass transport ($K_{bulk} = 0$) in Fig. 4a and bulk transport $K_{bulk} > 0$ with impermeable steps ($p = 0$) in Fig. 4b. For the first case, the best fit parameter values are $D_s/K_d = 100$ μm, $p/K_d = 2000$, and $g = 0.00819$, indicative of detachment-limited decay kinetics with highly permeable steps. In the second case, we obtain $D_s/K_d = 0.5$ μm, $K_{bulk}/K_d = 2.5$, and $g = 0.00354$. It is important to note that calculated curves obtained with bulk diffusion as the sole mass transport mechanism, e.g. with $D_s = 0$, are not consistent with the experimental results at any annealing temperature, T =

1550-1700 K, suggesting that the observed decay of TiN(111) islands *requires* the presence of surface mass transport.

Table I summarizes the materials parameters used to obtain the best fit solutions to the experimental data acquired at all four annealing temperatures in the two limits for which mass is not conserved locally: step permeability and bulk mass transport. Excluding bulk diffusion leads to high step permeabilities with $D_s/K_d$ values which are much larger than the average terrace width (~ 1000 Å). This is a signature of detachment-limited kinetics and, as such, is in agreement with previous high-temperature (T = 1000-1250 K) scanning tunneling microscopy (STM) measurements of 2D TiN island coarsening/decay kinetics on TiN(111) terraces [6]. If we include bulk mass transport with impermeable steps, we obtain $K_{bulk}/K_d$ ratios of order unity except at T = 1590 K, where $K_{bulk}/K_d = 10$. The calculated $D_s/K_d$ values in Table I also correspond to detachment-limited kinetics, but they are significantly smaller than the results obtained for permeable steps. The large difference between $D_s/K_d$ values in the permeable-step and bulk-exchange cases can be understood as follows. Step permeability by itself does not facilitate mass transport. It must be accompanied by fast surface diffusion in order to significantly violate local mass conservation. For bulk diffusion, this is not the case and local mass conservation is contravened even with a modest value of $D_s/K_d$.

Table I shows that $g$ increases monotonically with T and that the results are not very sensitive to whether local mass conservation is violated by step permeability or bulk mass exchange. Using the $g$ values in Table I and assuming, based upon STM measurements reported in Ref. [3], that the parameter $B = 0.23$ eV/Å is temperature-independent over the range of the present experiments, we calculate the step-step interaction strengths, $G = \left(\dfrac{\Omega}{kT}\right)^2 B^3 g$, to range from 0.03 eV-Å at T = 1559 K to 0.76 eV-Å at T = 1651 K. These values are comparable to results obtained for other materials including Si, Pb, and Cu [30].

Overall, we find that the agreement between the experimental data and the calculated results obtained with non-zero $K_{bulk}$ values is better than that obtained with high $p/K_d$ values. However, the differences are small and we cannot quantitatively distinguish between the two processes. Since step permeability, unlike bulk-diffusion, is a surface process and given that bulk-point defects usually have larger formation and diffusion energies than surface adspecies, the energetics controlling island decay should provide additional insights into the controlling mechanism. To this purpose, we measured island decay rates as a function of temperature.

The temperature dependence of the upper-island area decay rates $dA_1/dt$ is shown in Fig. 5. The data set includes 23 islands at four different temperatures between 1550 and 1700 K. From least-squares analyses of the results, we obtain an activation energy $E_d$ of 2.8±0.3 eV with a prefactor of $10^{12.9\pm0.8}$ Å$^2$-s$^{-1}$ for the decay kinetics of 2D TiN(111) islands stacked in 3D mounds. This is consistent with the previously reported value of 2.3±0.6 eV determined from STM observations of the decay of large 2D TiN(111) islands on atomically-flat terraces [6]. The fact that we obtain an $E_d$ value which is significantly lower than the bulk-mass transport barrier, 4.5±0.2 eV, measured for TiN(111) spiral step growth [23] provides further evidence that the dominant mass transport mechanism is surface, rather than, bulk diffusion. Thus, we attribute the decay of 2D TiN(111) islands in 3D concentric mound structures primarily to the presence of highly permeable steps in the detachment-limited regime. However, we note that the calculated step permeability rates are rather high which suggests that bulk mass transport is also playing a role. Since TiN(111) is known to have a wide single-phase field and can sustain both high anion and cation vacancy concentrations [31], mass exchange between the bulk and the surface cannot be ruled out. In the presence of a non-equilibrium point defect concentration in the bulk, dA/dt is expected to vary with annealing time and $N_2$ overpressure [32]. Thus, additional experiments, carried out over extended annealing times, investigating the effects of varying $N_2$ partial pressures are necessary to quantitatively determine the time-dependence of dA/dt, and hence the contribution due to bulk mass transport.

## 4. Summary

In situ high-temperature (T = 1550-1700 K) LEEM was used to study interlayer mass transport kinetics during annealing of 3D TiN(111) mounds consisting of stacked 2D islands. At each T, the islands decay at a constant rate, irrespective of their initial position in the mounds, indicating that mass is not conserved locally. We model the process based upon step flow, while accounting for step-step interactions, step permeability, and bulk mass transport. The results show that TiN(111) steps are highly permeable and exhibit strong repulsive temperature-dependent step-step interactions that vary from 0.03 eV-Å at 1559 K and 0.76 eV-Å at 1651 K. From temperature-dependent island decay rates, we obtain an activation energy of 2.8±0.3 eV, which is consistent with the detachment-limited decay of 2D TiN islands on TiN(111) terraces [6] and significantly lower than the bulk-mass transport barrier, 4.5±0.2 eV, measured for TiN(111) spiral step growth [23]. Based upon these results, we suggest that the rate-limiting process controlling the TiN(111) mound is surface, rather than bulk, diffusion in the detachment-limited regime.

**Acknowledgements**

The authors gratefully acknowledge the financial support of the U.S. Department of Energy (DOE), Division of Materials Science, under Contract No. DEFG02-91ER45439 through the University of Illinois Frederick Seitz Materials Research Laboratory (FS-MRL). We also appreciate the use of the facilities in the Center for Microanalysis of Materials, partially supported by DOE, at the FS-MRL.

**List of tables**

Table I. Materials parameters which provide the best agreement between the experimental data and the model describing the observed TiN(111) 3D mound decay behavior.

| T | Permeable steps ($K_{bulk} = 0$) | | | | Bulk exchange ($p = 0$) | | | |
|---|---|---|---|---|---|---|---|---|
| | $D_s/K_d$ (μm) | $p/K_d$ | g | G (eV-Å) | $D_s/K_d$ (μm) | $K_{bulk}/K_d$ | g | G (eV-Å) |
| 1559 K | 100 | 2000 | $9.78 \times 10^{-4}$ | 0.034 | 5 | 2 | $1.02 \times 10^{-3}$ | 0.036 |
| 1590 K | 200 | 2000 | $2.82 \times 10^{-3}$ | 0.095 | 0.5 | 10 | $< 10^{-5}$ | $< 3.4 \times 10^{-4}$ |
| 1622 K | 100 | 2000 | $8.19 \times 10^{-3}$ | 0.264 | 0.5 | 2.5 | $3.54 \times 10^{-3}$ | 0.114 |
| 1651 K | 200 | 2000 | $2.45 \times 10^{-2}$ | 0.7634 | 0.5 | 5.5 | $2.31 \times 10^{-2}$ | 0.7198 |

**List of figures**

Fig. 1. Representative BF-LEEM images (fields of view = 8.0 μm) of the surface morphologies of: (a) a sputter-deposited heteroepitaxial TiN(111) layer, and (b) and (c) different regions of the same sample following two deposition/annealing cycles. Figs. 1a, 1b, and 1c were acquired at electron energies $E_i$ of 13, 8, and 5 eV, respectively.

Fig. 2. (a)-(c): Representative BF-LEEM images (fields of view = 4.0 μm; $E_i$ = 15 eV) of a TiN(111) sample during annealing at T = 1552 K for times t of (a) 0, (b) 95, (c) and 230 s. (d) Areas A of islands labeled 1, 2, 3, and 4 in Figs. 2a-c plotted as a function of annealing time t.

Fig. 3. (a) Typical BF-LEEM image (field of view = 4.0 μm; $E_i$ = 13 eV) of a 3D conical stack of 2D homoepitaxial islands on TiN(111) during annealing at T = 1559 K. (b) Higher-resolution image (field of view = 1.7 μm) of the highlighted region in Fig. 3a. (c)-(f): island area A vs. annealing time t plots for four or more successive layers in the region highlighted in Fig. 3b at temperatures T of (c) 1559 K, (d) 1590 K, (e) 1622 K, and (f) 1651 K.

Fig. 4. (a), (b): Plots of island radii $r_i$ vs. annealing time t at T = 1622 K for the four samples corresponding to the data in Fig. 3e. The dashed and solid lines in Figs. 4a and 4b are calculated curves describing the LEEM data (open circles). The dashed lines in Fig. 4a are obtained assuming that the steps are impermeable and that there is no net bulk mass transport, i.e. $p = K_{bulk} = 0$. The solid lines in Figs. 4a and 4b are obtained assuming permeable steps ($p > 0$) with no bulk mass transport ($K_{bulk} = 0$) and bulk transport ($K_{bulk} > 0$) with impermeable steps ($p = 0$), respectively.

Fig. 5. A plot of 2D TiN(111) upper island decay rates $dA_1/dt$ vs. annealing temperature T. The solid line is a least-squares fit to the data and corresponds to an activation energy $E_d$ of $2.8\pm0.3$ eV with a prefactor of $10^{12.9\pm0.8}$ $\text{Å}^2\text{-s}^{-1}$.

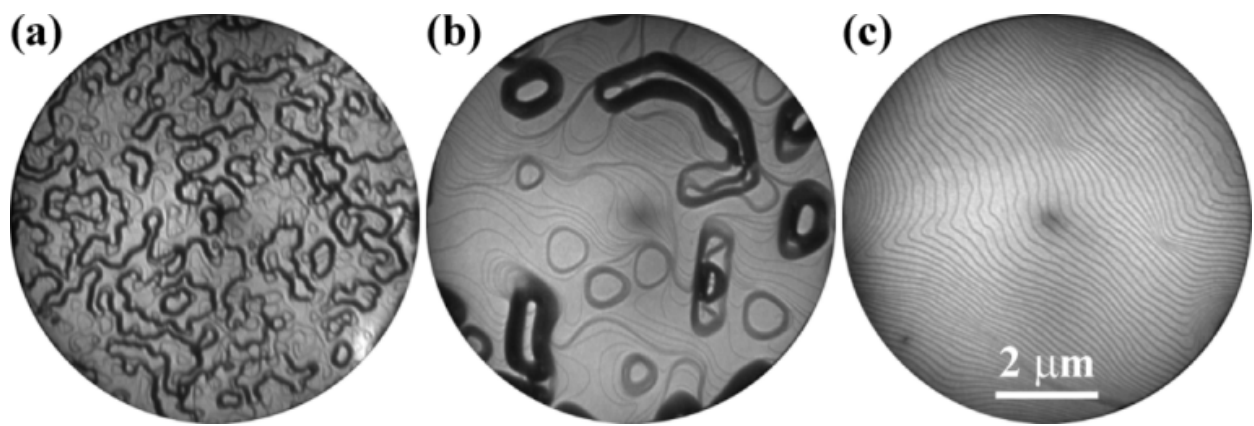

Fig. 1. S. Kodambaka et al.

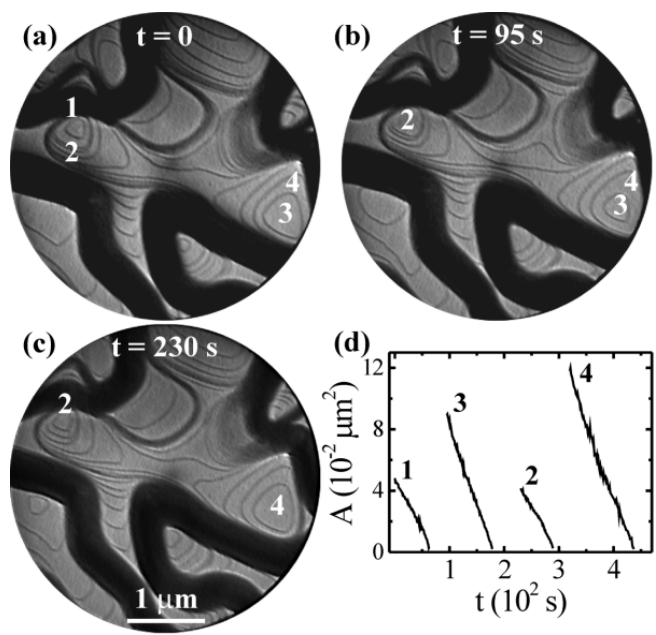

Fig. 2. S. Kodambaka et al.

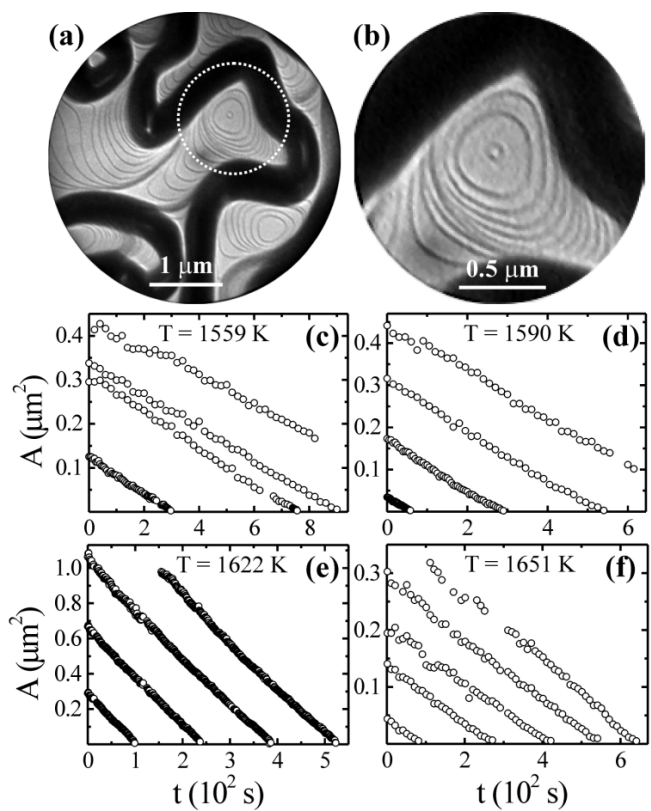

Fig. 3. S. Kodambaka et al.

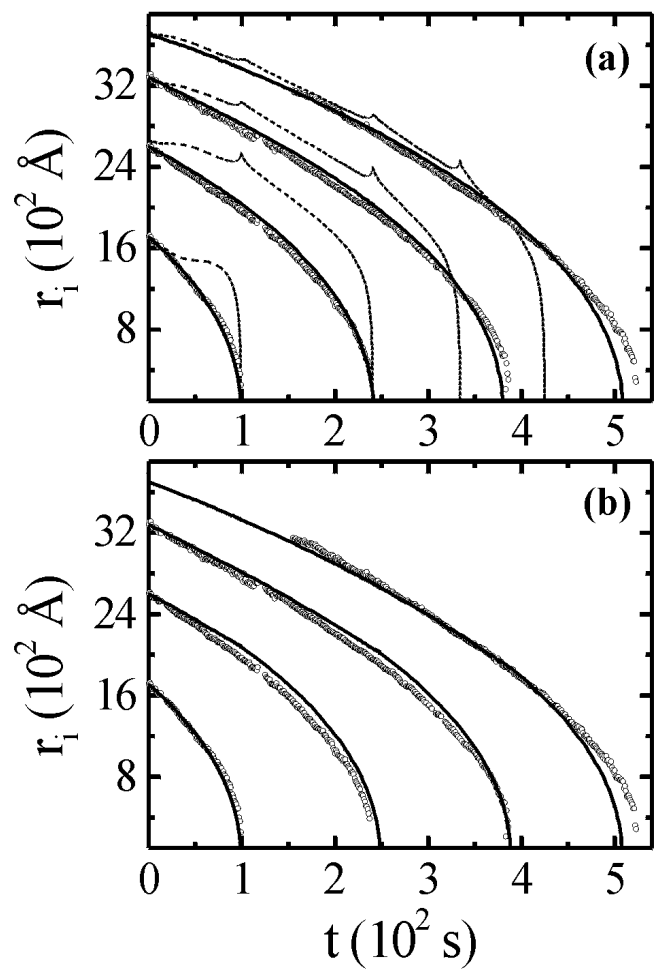

Fig. 4. S. Kodambaka et al.

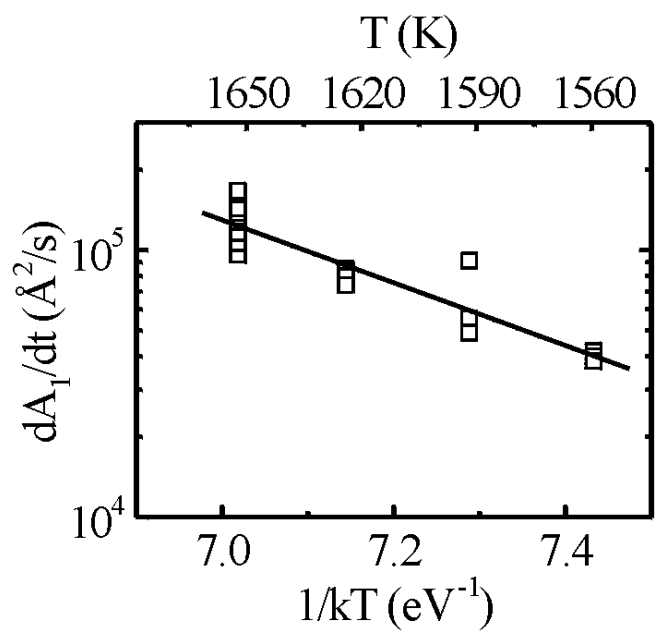

Fig. 5. S. Kodambaka et al.